%% file: main.tex
\documentclass[sigconf, prologue,table]{acmart}

\acmConference[ICSE 2024]{46th International Conference on Software Engineering}{April 2024}{Lisbon, Portugal}

\AtBeginDocument{%
  \providecommand\BibTeX{{%
    \normalfont B\kern-0.5em{\scshape i\kern-0.25em b}\kern-0.8em\TeX}}}

\usepackage[framemethod=tikz]{mdframed} % 
\usepackage{enumitem}
\usepackage{soul}
\usepackage{stfloats}

\usepackage{soul}
\usepackage{colortbl}
\usepackage{adjustbox}
\usepackage{caption}
\usepackage{multirow}
\usepackage{tikz}
\usepackage{xcolor}
\usepackage{refcount}
\usepackage{hyperref}
\usepackage{tabularx}
\usepackage{threeparttable}
\usepackage{multicol}
\usepackage{float}

\usepackage[normalem]{ulem}
\newif\ifdraft
\draftfalse %hides boldifications and comments
\newif\ifrevising
   \revisingfalse %makes revisions normal appearance
\newcommand{\revised}[1]{{\ifrevising{\color{blue}#1}\else{\color{black}#1}\fi}}
 \newcommand{\deleted}[1]{{\ifrevising{\relax}\else\relax\fi}}

\begin{document}

%title{Design guidelines for a conversational agent}

\title{How Far Are We? The Triumphs and Trials of Generative AI in Learning Software Engineering}
% \title{How Far Are We? The Triumphs and Trials of Generative AI in Supporting Students in Software Engineering}
% ...in Supporting Students in Software Engineering

\author{Rudrajit Choudhuri}
\affiliation{%
  \institution{Oregon State University}
  \city{Corvallis}
  \state{OR}
  \country{USA}
}
\email{choudhru@oregonstate.edu}

\author{Dylan Liu}
\affiliation{%
  \institution{Oregon State University}
  \city{Corvallis}
  \state{OR}
  \country{USA}
}
\email{liudy@oregonstate.edu}

\author{Igor Steinmacher}
\affiliation{%
  \institution{Northern Arizona University}
  \city{Flagstaff}
  \state{AZ}
  \country{USA}
}
\email{igor.steinmacher@nau.edu}

\author{Marco Gerosa}
\affiliation{%
  \institution{Northern Arizona University}
  \city{Flagstaff}
  \state{AZ}
  \country{USA}
}
\email{marco.gerosa@nau.edu}

\author{Anita Sarma}
\affiliation{%
  \institution{Oregon State University}
  \city{Corvallis}
  \state{OR}
  \country{USA}
}
\email{anita.sarma@oregonstate.edu}

%
% By default, the full list of authors will be used in the page
% headers. Often, this list is too long, and will overlap
% other information printed in the page headers. This command allows
% the author to define a more concise list
% of authors' names for this purpose.
\renewcommand{\shortauthors}{Choudhuri et al.}

\newcommand{\explaintwo}[1]{%
\par%
\noindent\fbox{%
    \parbox{\dimexpr\linewidth-2\fboxsep-2\fboxrule}{#1}%
}%
}

% \newcounter{guidelineno}
%\newcommand{\guideline}[1]{\noindent\textcolor{blue}{\textbf{DG \refstepcounter{guidelineno}\theguidelineno:} #1}}

%\newcounter{fullguidelineno}
% \newcommand{\fullguideline}[2]{\noindent\textcolor{blue}{\textbf{DG \refstepcounter{guidelineno}\theguidelineno: #1} #2}}
\newcommand{\blue}[1]{\textcolor{blue}{#1}}

\begin{abstract}
Conversational Generative AI (convo-genAI) is revolutionizing Software Engineering (SE) as engineers and academics embrace this technology in their work. However, there is a gap in understanding the current potential and pitfalls of this technology, \revised{specifically in supporting students in SE tasks}. In this work, we evaluate through a between-subjects study (N=22) the effectiveness of \revised{ChatGPT, a convo-genAI platform,} in assisting students in SE tasks. Our study did not find statistical differences in participants' productivity or self-efficacy when using \revised{ChatGPT} as compared to traditional resources, but we found significantly increased frustration levels. Our study also revealed 5 distinct faults arising from violations of Human-AI interaction guidelines, which led to 7 different (negative) consequences on participants. 
%Through this study, we provide a comprehensive evaluation of convo-genAI's current potential and pitfalls while simultaneously informing future AI design.
\end{abstract}

\begin{CCSXML}
<ccs2012>
<concept>
<concept_id>10003120.10003121.10011748</concept_id>
<concept_desc>Human-centered computing~Empirical studies in HCI</concept_desc>
<concept_significance>500</concept_significance>
</concept>
</ccs2012>
\end{CCSXML}

\ccsdesc[500]{Human-centered computing~Empirical studies in HCI}
\keywords{Empirical Study, Software Engineering, Generative AI, ChatGPT}

\maketitle

\input{sec/1.intro}

\vspace{-3mm}
\input{sec/2.Method-new}
\vspace{-3mm}
\input{sec/3.Results}

\vspace{-2mm}
\input{sec/4.Discussion}
\vspace{-2mm}
\input{sec/5.Related-Work}
\vspace{-1mm}
\input{sec/6.Threat}
\vspace{-3mm}
\input{sec/7.conclusion}


\begin{acks}
We thank Samarendra Hedaoo for his insights in the meetings about the SE course. We thank all participants who took part in the study for their time and effort. We thank members of the EPIC lab at the university:
Amreeta Chatterjee,  Emily Garcia, Mariam Guizani, Zixuan Feng, and Bianca Trinkenreich, 
among others, for their support and valuable feedback, and Andrew Anderson for his early insights. 
This work was partially supported by the National Science Foundation under Grant Numbers: 2235601, 2236198, 2247929, 2303042, and 2303043. Any opinions, findings, conclusions, or recommendations expressed in this material are those of the authors and do not necessarily reflect the views of the sponsors.
\end{acks}

\bibliographystyle{ACM-Reference-Format}
\bibliography{acmart}

\appendix

\end{document}

%% file: sec/1.intro.tex
\section{Introduction}
\label{sec:intro}
% Story:
% **Convo-genAI is all the craze, including ed....
% ** Past work has shown
% 1) how convo agents are helpful
% 2) how genAI is helpful
% ** But all of this is for intro CS, what about higher division courses
% ** SE is one such, has more complex...convo+genAI tools can be helpful in giving context help
% ** Thus, we do blah
% **

%%%\boldification{convo GenAI discussions go both ways, we dont know how SE ed can leverage GenAI.}
The advent of Conversational Generative AI (convo-genAI) is proving to be a ``Gutenberg moment'' across education and business, and software engineering is no exception. Convo-genAI systems (e.g., ChatGPT~\cite{GPT4}, Google Bard~\cite{bard}, Meta LLaMA~\cite{llama}) generate novel content, be it a haiku or a relevant code snippet, informed by pre-existing data and minimal input. Additionally, these tools provide a conversational interface, enabling users to interact using natural language to generate outputs fine-tuned to their needs. Discussions on the use of generative AI (genAI) range from how it signals the ``end of programming''~\cite{welsh2022end, yellin2023premature} to how it can transform software engineering for the better~\cite{daun2023chatgpt, ma2023scope}. 

%\boldification{Introductory CS education has investigated conversational agents and GenAI}
Given the nascency of this innovation, it remains unclear how it can be leveraged in education, and uncertainty overshadows its potential benefits.
The use of both conversational agents and genAI has been investigated in the context of Introductory Computer Science (CS).
% Conversational interfaces allow users to dialog in natural language have been studied in the context of CS education. 
Prior work on conversational agents has demonstrated their effectiveness for non-specialists and learners, as they facilitate dialogue with an "expert" \cite{valtolina2020communicability, walgama2017chatbots, grudin2019chatbots, garcia2023support}. For example, Loksa et al. \cite{loksa2016programming} found that high-schoolers in an introductory programming summer camp benefited when they could seek help by articulating their problem-solving strategies and their current task state.  
Further, the new generation of students prefers talking with chatbots over talking to a person~\cite{alton2017phone}. Indeed, there has been extensive work on conversational agents for education~\cite{Okonkwo2021chatbots, wollny2021we} focusing on students' engagement \cite{bii2013chatbot,fryer2017stimulating}, self-directed learning \cite{pereira2016leveraging}, and tutoring~\cite{tamayo2017designing}. 
GenAI systems have also been explored in this context \cite{denny2022robosourcing, macneil2022generating, verleger2018pilot}. %as well as a pair programming buddy for  developers\cite{bird2022taking}. 
% \citet{verleger2018pilot} investigated the use of AI-driven chatbot to [result] use in problem solving. Similarly, \citet{macneil2022generating} looked into how these systems can be leveraged for generating code explanations. 
But these studies have either been focused on using genAI to solve algorithmic problems~\cite{peng2023impact,sun2022investigating} (i.e., programming) or on improving genAI technology and output~\cite{li2023blip,white2023prompt} (i.e., better AI).

%\boldification{SE is well suited for multiple reasons, but thus far, no one has looked at genAI in SE courses, which are particularly suited for this}
Currently, there is a gap in understanding how convo-genAI systems can be leveraged for learning advanced CS topics, such as Software Engineering (SE), where the learning objectives include unique complexities and contextual decision-making involving subjectivity and multiple trade-offs \cite{pinto2017training, pinto2019training}. In fact, SE education extends beyond the confines of classroom education, into the workplace, where software developers need to learn job-relevant topics, processes, and tools~\cite{begel2008novice}.
Learning in these situations requires contextualized assistance.
Given that convo-genAI systems are capable of providing such assistance, coupled with natural language dialogue capabilities, they can be particularly well suited for SE topics~\cite{daun2023chatgpt}, where traditional chatbots have fallen short so far~\cite{farooq2016human}).
% , dominic2020conversational}). 
%Therefore, it is essential to have a clear understanding of where and to what extent these systems can be helpful in software engineering education. 

Thus, in this paper, we investigate:
\textbf{\textit{(RQ1): How effective is convo-genAI in helping students in software engineering tasks?}} 
%Specifically we investigated how well it is in assisting students with SE concepts (fixing code functionalities, good coding practices), processes and tools.
%
We answered this research question through a between-subjects user study, where the Experimental group solely used ChatGPT, while participants in the Control group could use any resource other than genAI tools.
We selected ChatGPT as a representative convo-genAI as it was state-of-the-art when writing this paper. We recruited 22 students enrolled in undergraduate software engineering courses at our university.
Participants in both groups completed three SE tasks (fixed code functionalities, removed code smells, and contributed to a GitHub repository).  
We found no statistical differences between the two treatments in terms of participants' task performance or overall self-efficacy, but using ChatGPT increased participants' frustration.

\begin{figure*}[hbt]
\centering
\vspace{-10px}
\includegraphics[width=0.85\textwidth]{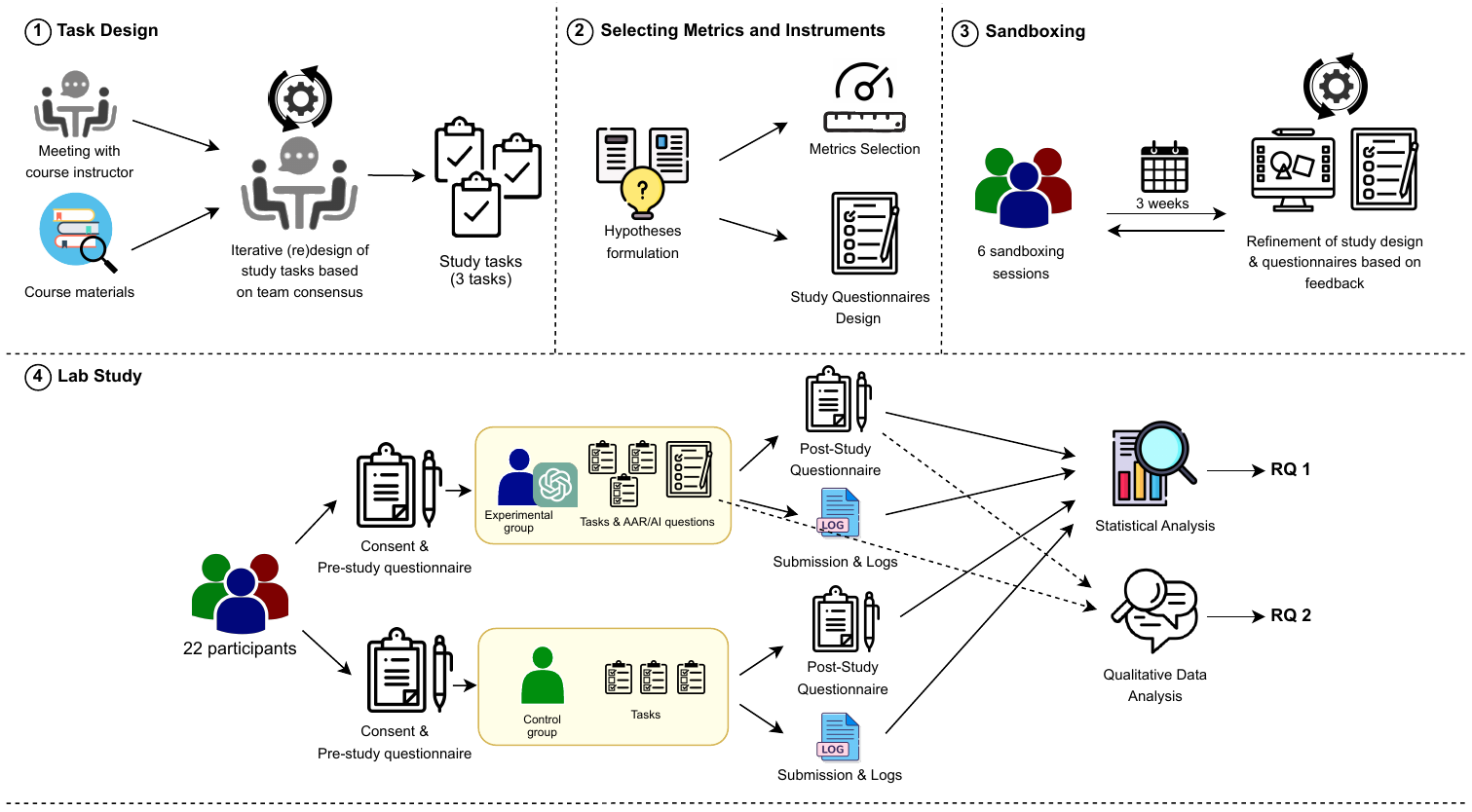}
\vspace{-12px}
\caption{Overview of the research design} 
\vspace{-12px}
\label{fig:design}
\end{figure*}

Furthermore, to gain a comprehensive understanding of how convo-genAI can be effectively used and to inform its future design, it is essential to identify and evaluate its current pitfalls. This leads to our second research question:
\textit{\textbf{(RQ2): What are the current pitfalls in convo-genAI?}} 
Specifically, we investigated the faults that \revised{ChatGPT}
currently makes \revised{in the context of helping students in SE tasks}, the causes of these faults, and their consequences for the \revised{participants}.
To answer this question, we employed After-Action Review for AI (AAR/AI)~\cite{dodge-aar}, a recent AI assessment process that allows end users to gauge AI faults~\cite{khanna2022finding} through a set of questions. The purpose of using AAR/AI was to assess the faults that participants in the Experimental group could perceive during their interaction with ChatGPT and its consequences for the participants. Qualitative analysis of these responses revealed \textit{five fault categories and seven consequences stemming from these faults}. 

Additionally, we used Microsoft's design guidelines for Human-AI Interaction (HAI)~\cite{amershi2019guidelines} as a lens to assess the ties between these faults and the violation of specific guidelines. Since these guidelines impact how users interact with the tool~\cite{li2022assessing}, it is important to get their perceptions of its violations. %Thus, we collected (Experimental group) participants' feedback on their perceptions of whether any of the HAI Guidelines were violated in a post-study Likert-scale questionnaire. 
Doing so adds the benefit of participants reflecting on the tool and their interactions, thereby promoting metacognition~\cite{nelson1994investigate}. A majority of participants reported that ChatGPT: (1) did not clearly outline its capabilities and limitations, (2) did not support efficient correction, (3) did not scope itself when in doubt, and (4) lacked transparency in its decision-making process. Our analysis revealed that these guideline violations were the root cause of the faults.

%\rudy{need sunset} 
%\boldification{Closing statement}
The primary contribution of this paper is an evaluation of the potential benefits and challenges of providing a convo-genAI system, specifically ChatGPT, to assist in software engineering tasks. Convo-genAI's ability to generate contextualized help along with natural language dialog capabilities can revolutionize software engineering. However, the inconsistency in its behavior and output pose significant challenges to its use in educational settings.  
Such systems, in their current state, can lead to user frustration and cognitive overload, which can be discouraging especially for novices, undermining their self-efficacy and potentially triggering an early exit from the field. 
Future design should thus incorporate the insights related to the current pitfalls and guideline violations to design for better user interactions. Educators, students, and self-learning professionals should also be aware of the consequences of using these tools to acquire new knowledge and skills. 

%(b) be inclusive of different interaction styles~\cite{anderson2021diverse}.

%% file: sec/2.Method-new.tex
\section{Method}
\label{sec:method}

%% \boldification{The proposed work follows a multi-step approach to answer our RQs. We conducted controlled studies with SE students}
%%\boldification{We implemented a between subjects design (CG vs non-LLM)}
Our study goal is to understand the effectiveness and limitations of Conversational Generative AI (convo-genAI) compared to traditional online resources in helping software engineering students. To achieve this goal, we conducted a controlled experiment with students enrolled in undergraduate-level software engineering courses. Our study employed a between-subjects design, dividing the participants into two groups: an Experimental group that exclusively used ChatGPT 
% (a convo-genAI platform)
\revised{(GPT-4)}; 
and a Control group that could use any online resources except genAI tools. We designed tasks based on in-depth discussions with the course instructor and careful analyses of the course materials. We measured multiple outcomes to comprehensively understand the effects and limitations of convo-genAI. The study protocol was refined through sandboxing sessions. After that, we conducted the experiment, as depicted in Figure \ref{fig:design}.% provides an overview of the research design, which we detail in the following sections.
%, we employed different metrics and questionnaires to assess multiple perspectives. Additionally, we iteratively refined the study design through sandboxing sessions, making necessary adjustments based on the feedback and discussions.  Following this iterative refinement process, we recruited our actual study participants and conducted the study in person at the university lab. 
%%\boldification{The initial phase of the study design was focused on crafting suitable tasks. Then we focused on structuring our study and employed multiple metrics for data collection}
%%\boldification{We then refined our design through sandboxing making amendments as required.}
%The initial phase of the study design involved crafting suitable tasks by drawing insights from course materials (syllabus and educational content) and in-depth discussions with the course instructor. Therefore, we focused on structuring our study to triangulate information and collect data using multiple metrics and instruments. Subsequently, we carried out six sandboxing sessions with students, making necessary amendments to the study design as required. These modifications were implemented following discussions and consensus within the research team. 
\vspace{-3.5mm}
\subsection{Task Design}

%% \boldification{Step 1: Task design: To achieve that, we had a meeting with a course instructor and identified what our participants were (un)familiar with.} 
To design appropriate tasks, we first investigated students' current backgrounds and the skills they would learn in the course. Three researchers analyzed the course documentation and had in-depth discussions with the instructor \revised{about the learning objectives of the course}. From the discussion, we learned that the students had low to medium familiarity with git and GitHub and that Python was the programming language used in the course. 
Furthermore, it was understood that the students were novices in software engineering and had a rudimentary understanding of good programming styles, API usage, and web scraping. We also learned that the students had not yet been formally introduced to the concept of code smells~\cite{fowler1997refactoring}. %\cite{codesmells}, which is an important part of cultivating good programming practices \cite{fowler1997refactoring}. 
% We explored these topics in our tasks.
\revised{We tailored our tasks to align with the topics the instructor emphasized as key learning objectives in the course. Specifically, we focused on four topics: programmatic API usage, debugging, code quality, and version management, through the following 3 tasks: (1) fixing code functionalities involving third-party APIs, (2) removing code smells, and (3) contributing changes to a repository via a pull request.}

%%\boldification{ After the initial discussion, the team came up with the study tasks (what are they) and iteratively revised it. The tasks were of moderate complexity (why)}
The tasks were designed to be moderately challenging for the students. The research team held multiple meetings over two weeks to define and review the tasks. 
We then created a small Python program and hosted it in a GitHub repository\footnote{See supplemental \cite{supplemental2023}: \url{https://zenodo.org/record/8193821}}. 
%It had three tasks: (1) fixing code functionalities, (2) removing code smells, and (3) contributing changes to the repository via a Pull Request (PR). 
%We selected and sequenced these tasks to mimic a typical contribution scenario that involves software engineers creating their development branch, rectifying the code, improving code quality, and submitting their changes to the project. 
We intentionally used a small program to manage task complexity. Additionally, we disorganized the code structure, added poorly crafted comments, and used non-intuitive function names to simulate common software engineering challenges. We conducted multiple sandboxing sessions with participants who had low to medium familiarity with software engineering, Git, and GitHub to validate the appropriateness and the level of complexity of these tasks: 

%%\boldification{ The first task was ….}
\textit{Task 1:} Participants needed to \textit{create their own git branch} and \textit{debug} code functionalities in their branch (3 subtasks). They were restricted from modifying some parts of the program or changing the test instances, ensuring that the only way to pass the validation was by fixing some code functionalities, as follows: (1) Participants had to correct the `check\_palindrome' function (logical programming exercise) that checked whether a number was a palindrome. We reversed the logic of this function, creating a bug. (2) Participants were asked to fix the `check\_weather' function (API usage exercise). The objective of the function was to retrieve weather data using the NOAA\_SDK library\footnote{https://pypi.org/project/noaa-sdk/} in Python, given specific coordinates. The correct method for this operation is to utilize the `points\_forecast' method from the API. We introduced a bug by using the `get\_forecasts' method, which actually takes the postal code and country as parameters, and passed latitude and longitude values instead. (3) Participants were tasked with fixing a function named `check\_webpage' (basic web scraping). The purpose of this function was to leverage the Selenium library\footnote{\url{https://pypi.org/project/selenium/}} in Python to capture all visible elements from a webpage.
%Interestingly, we used an obsolete code snippet for this function that was originally generated by ChatGPT.

%%\boldification{The second task was …}
\textit{Task 2:} The second task focused on improving \textit{code quality by identifying and removing code smells}. 
%Participants were tasked with detecting occurrence of smells within the code, and implementing appropriate fixes.
% We intended to familiarize participants with the concept of code smells, enabling them to detect their occurrence within the code and implement appropriate fixes, and this motivated our task design. 
We introduced these code smells: global variables, unused imports, dead code (commented-out code), and magic numbers (constants without context).

%%\boldification{ Finally, the third task was….}
\textit{Task 3:} The third task was designed to familiarize students with the \textit{configuration management process} by contributing to a repository using git/GitHub. Participants were required to commit their contributions to the remote branch and create a pull request (PR) to the base branch. To maintain the integrity of the main branch, we instituted branch protection rules to prevent modifications in the base branch.

%%% POTENTIAL THREAT: Students knew beforehand the topics

\vspace{-2.5mm}
\subsection{RQs, Metrics and Instruments}
\label{subsubsec: instruments}
%%\boldification{ Step 2: Next  we employed multiple metrics and instruments to concrete our understanding of students’ perceptions of the technology and the effects of interacting with it. }

We employed multiple metrics and instruments (Table \ref{table:metrics}) to understand students' perceptions of ChatGPT and the effects of interacting with it. In the following sections, we detail each RQ and the instruments used to answer them.

\begin{table}[h]
\vspace{-6pt}
 \caption{\small{Metrics and instruments and their relation to the RQs}} % title of Table
 \label{table:metrics} % is used to refer this table in the text
 \vspace{-8pt}
% \centering
\input{table/Metrics-Instruments}
\vspace{-8pt}
\end{table}
\subsubsection{RQ1: How effective is convo-genAI in helping students in software engineering tasks?}
Toward answering RQ1, we came up with three hypotheses and assessed the \revised{participants'} intention to continue using \revised{ChatGPT}, as detailed below.

% We used these instruments to evaluate \textit{"how far are we?"} with Generative AI supported software engineering education.
%%\boldification{For RQ1, we had the following hypotheses.} 

\noindent{\textit{\textbf{H1: Students using ChatGPT for the study tasks perceive lower cognitive load than students using alternate resources.}}}

Cognitive load is an important factor in designing effective scaffolding tools as it influences effective learning and end-user productivity \cite{abbasi2019effect}. Previous literature has shown that interacting with conversational agents (chatbots) significantly reduces cognitive load for end users in certain contexts \cite{schmidhuber2021cognitive, abbasi2019effect, brachten2020ability}. Therefore, as a first step, we evaluated convo-genAI's impact on participants’ perception of cognitive load in our context. \textit{We hypothesized that students using ChatGPT for the study tasks (Experimental group) perceive a lower cognitive load compared to students who use alternate resources (Control group).} We used the original NASA Task Load Index (TLX)\cite{hart1988development}---a validated and widely used post-study instrument---for measuring cognitive load across six dimensions (mental, physical, and temporal demand, performance, effort, and frustration). %We used 21-point Likert scale (aligned with the original scale).% extending from 'very low' to 'very high.'

\noindent{\textit{\textbf{H2: ChatGPT positively impacts students' productivity.}}}

\citet{schmidhuber2021cognitive} revealed that chatbots can positively impact end users' productivity (measured in average correctness and average time required). Therefore, in our context, \textit{we hypothesized that ChatGPT positively impacts students' productivity} and employed the variables from Xu et al.'s study~\cite{xu2022ide} that assessed productivity in terms of task performance: (a) task correctness and (b) time to completion. However, we excluded (b) since we time-boxed the tasks (see Section \ref{sec:sandbox}) and participants in both groups utilized all of the allotted time for each task.
% Therefore, it is expected that if a conversational agent is useful, end users can complete the tasks faster without compromising the output quality. 
To analyze the task correctness, two researchers used rubrics for each task---which is a prevalent approach in computer science education research~\cite{buragga2013rubric, catete2017application}. The rubrics were collaboratively developed beforehand (and are detailed in the supplemental \cite{supplemental2023}). The assessment consisted of number of tests correctly passed, code smells removed, successfully committed contributions and pull requests opened \revised{(assessed based on participants' code submissions and their GitHub log data)}. %Task 1 and 2 had three points each, whereas Task 3 had two points in total. 
We followed blind grading for code solutions (not knowing whether it came from the Experimental/Control group) to reduce potential bias. 

\noindent{\textbf{\textit{H3: ChatGPT promotes students' self-efficacy.}}}

Self-efficacy reflects an individual's perceived ability to successfully accomplish a task and can influence one’s actual ability to complete a task~\cite{bandura1986explanatory}. Self-efficacy is considered a robust predictor of achievement~\cite{bandura1993perceived} and motivation~\cite{chang2014effects}. Recent studies have shown that conversational tools effectively promote students' self-efficacy \cite{chang2022promoting, park2023utilizing}. Hence, in our context, \textit{we hypothesized that ChatGPT promotes students' self-efficacy.} To capture this construct, we adapted a questionnaire from \citet{steinmacher2016overcoming} and aligned it with the context of our study (see supplemental \cite{supplemental2023}). We used a 5-point Likert scale (`Strongly Disagree' to `Strongly Agree') and administered the questionnaire in a before-after design, which allowed us to assess the impact of the resources on participants' self-perceived efficacy.

%%\boldification{Furthermore, continuance intention reflects satisfaction and positive perception of tool. we assessed continuance intention using a questionnaire inspired from previous work- CHI'23}

\vspace{0.9mm}
\noindent{\textbf{\textit{Users' continuance intention}}}: Prior research highlights that the users' continuance intention reflects their satisfaction and positive perception towards the tool \cite{ashfaq2020chatbot}. \citet{park2011antecedents} evaluated continuance intention using a direct likelihood question. Echoing this method, we assessed the participants' continuance intention by presenting them with three statements:  (1) Based on this experience, I plan to use ChatGPT to learn Software Engineering concepts; (2) Based on this experience, I do not plan to use ChatGPT to solve similar kinds of tasks; and (3) I would recommend ChatGPT to my friends if they need assistance with Software Engineering. The participants were asked to rate their level of agreement for each of these statements on a 5-point Likert scale (`Strongly Disagree' to `Strongly Agree').
%\footnote{We used 2 positive points and 2 negative for continuance intention. We used the 'undecided' option only for this question to account for the possibility that participants might be unsure about their future use of the tool for the specific context}. 
% ===> This undecided in the middle is a THREAT to validity. I'm removing from the paper.

\begin{table*}
    \footnotesize
    \vspace{-2mm}
    \caption{\label{tab:aar-ai-steps} \small{AAR/AI steps and our adaptations. The Empirical context column explains how we realized the method in our study. Steps 3 to 6 were “inner loop” questions we repeated for all three tasks.}}
     \vspace{-8pt}
\input{table/AARAI-table}
   \vspace{-12pt}
\end{table*}

%%\boldification{ RQ2 method: To assess any AI faults, we need a standardized method. We chose AAR/AI. (info shifted to Background)}
%\rudy{Don't change this specific paragraph unless something does not seem right.}
% \vspace{-mm}

\subsubsection{RQ2: What are the current pitfalls in convo-genAI?} To answer RQ2, we wanted \revised{participants} from the Experimental group to assess AI faults. A consistent evaluation of an AI tool requires a standardized assessment process to avoid users adopting ad-hoc approaches, which can result in variations when evaluating the same AI tool. Therefore, we employed the After-Action Review for AI (AAR/AI) instrument~\cite{dodge-aar}, a standardized AI assessment process designed to aid humans in effectively gauging AI faults~\cite{khanna2022finding}. \citet{khanna2022finding} provided empirical evidence that integrating AAR/AI can aid end users in uncovering a significantly larger number of faults with greater precision. %Drawing from \citet{dodge-aar}, participants using AAR/AI mentioned that it fostered logical reasoning about the AI agent and encouraged them to reason about the AI behavior from diverse perspectives. 
AAR/AI is a recent member of the After-Action Review (AAR) \cite{morrison1999foundations} family, devised by the U.S. military in the 1970s as a facilitated debriefing method. AAR has been used for decades and has been successfully adapted to different domains~\cite{ishak2017slides, sawyer2013adaptation}.  %davies2019enhancing, including transportation, healthcare, and emergency response.

% Most of the adaptations have been successful and a meta analysis of 61 studies revealed that incorporating the AAR method produces a practical moderate effect overall \cite{keiser2021meta}.
% A meta-analysis of 61 studies revealed that the AAR method produces a practical moderate effect overall \cite{keiser2021meta}.
%%\boldification{ Why did we care? Did studies show that humans assess AI better with AAR/AI?}
% AAR/AI is a form of retrospective analysis tailor-made for a human assessor to review AI’s behavior. 

%%\boldification{ How is AAR/AI done and how did we adapt it?}
The AAR/AI steps are derived from the ``DEBRIEF'' adaption by Sawyer and Deering \cite{sawyer2013adaptation}. There are 7 steps: \textbf{D}efining the rules, \textbf{E}xplaining the objectives of the AI agent, \textbf{B}enchmarking the performance of the agent, \textbf{R}eviewing what was supposed to happen, \textbf{I}dentifying what actually happened, \textbf{E}xamining why, and finally \textbf{F}ormalizing learning. AAR/AI is highly adaptable as it offers flexibility within each step of its process, accommodating customization to suit the specific needs of AI assessment in different domains. We adapted the AAR/AI steps as follows (Table 2):

\textit{\textbf{Defining the rules and explaining the objectives:}} Each session started with the researcher providing an overview of the interfaces that the participants would use (Git/GitHub, Visual Studio Code) and the study tasks. We then told them the purpose of the assessment: ChatGPT was expected to deliver contextual, unambiguous, and accurate information (Steps 1-2, Table \ref{tab:aar-ai-steps}). 
%In this briefing, the Experimental Group was explicitly instructed that ChatGPT would be their sole resource for the tasks.  

\textit{\textbf{The inner loop: What \& Why:}} Following task explanations and before each task initiation, participants answered, ``What do you think should happen when you use ChatGPT for this task?'' After each task, they responded to: ``What actually happened when you used ChatGPT for this task?'', ``Why do you think ChatGPT behaved this way?'' and ``To what extent did you modify ChatGPT’s responses for solving the task? Briefly explain why.'' (Steps 3-6, Table \ref{tab:aar-ai-steps}). The inner loop was repeated for all three tasks. The time to answer these questions was not counted towards task completion.
%not counted in the time to finish the tasks.
%To gain a more concrete understanding of the participants' actions in specific contexts (during each task), we modified Step 6 by substituting "would" with "did" (See Section \ref{sec:sandbox}). 

\textit{\textbf{Formalizing learning:}} Once all tasks were completed, we asked the participants: "What went well?", "What did not go well?", and "What could be done differently next time?" (Step 7, Table \ref{tab:aar-ai-steps}). 

%%%\boldification{ Next we needed to link the faults to some guideline violations. There are so many guidelines, but we chose Amershi’s ones (Why). We adapted it as statements as the guidelines were high level advice.} \rudy{Need justification here}

%%\as{Make guidelines new para coz its a contri in our paper - ...These guidelines for system design. We investigate when ChatGPT from this lens...to find where it violates these guidelines. How these violations led to effects on users to understand future improvements.}

To inform future design, assessing why the faults occur is important. Human-AI interaction guidelines inform AI system design and operation and can be used to understand where AI systems fail. Wright et al.'s~\cite{wright2020comparative} exploration of guidelines from three large tech companies (Apple, Google, and Microsoft) offered over 200 guidelines, ranging from initial considerations of AI to the curation of the models, the deployment of the AI-powered system, and the human-AI interface. Notably, Wright et al. found that while both Apple's~\cite{AppleGuidelines} and Google's~\cite{GoogleGuidelines} guidelines were created with the developer in mind, Microsoft's guidelines~\cite{amershi2019guidelines} were designed with a keen focus on the user. 

Therefore, we used Microsoft's guidelines~\cite{amershi2019guidelines} as our lens to examine whether ChatGPT's faults stemmed from potential guideline violations (the causes). Following this, we analyzed the effect of these faults on the \revised{participants} (the consequences). We adapted the general recommendations to our context, and participants rated a set of statements using a 5-point Likert scale (`Strongly Disagree' to `Strongly Agree'). These adaptations were made after team discussions and were refined during our sandboxing sessions.

%%\boldification{Step 3: After the initial study design, it is important to test whether the design helps us measure what we want. We did sandboxing. We changed (what) and (why).}
%\vspace{-4mm}
\subsection{Sandboxing} 
\label{sec:sandbox}
We conducted 6 sandbox sessions with CS students. We conducted the first two sessions via Zoom, and it became apparent that this setup posed several challenges (long setup times, limitations in our ability to control the environment, and difficulties in library installations). We decided to transition to an in-person lab setting. The sessions were planned to take 2 hours, but due to participant fatigue, we adjusted them to last 80 minutes and time-boxed the tasks: Task 1 was allotted 20 minutes, and tasks 2 and 3 had 10 minutes each. This change facilitated on-time conclusions and time for briefing and questionnaires (40 minutes).

% \boldification{ We changed AAR/AI steps a bit based on feedback and observations. Additionally, we reverse coded some items to avoid acquiescence bias. Lastly, we also dropped two guidelines as they were not relevant to our study.}
Our sandboxing also revealed difficulties with the AAR/AI process: participants found it burdensome, resulting in sparse responses. Echoing a similar concern raised by~\citet{dodge-aar}, we revised steps 3, 4, and 6---employing a mix of open and closed-ended questions---and adjusted the wording of the questions to improve their clarity. Additionally, we reverse-coded some items with negative connotations (low scores indicating agreement) in the Human-AI interaction guideline statements to counter acquiescence bias \cite{baumgartner2001response}, a tendency where participants agree with statements regardless of their content or due to laziness, indifference, or automatic accommodation to a response pattern. Lastly, all the researchers agreed to omit Guideline 3 (time services based on context) due to its irrelevance in our context (ChatGPT remains idle unless prompted). Guideline 18 (notify users about changes) was also dropped, given its lack of applicability, as there were no major ChatGPT updates during the study's timeline (May 15 - June 2, 2023).

%In particular, Step 6 underwent a substantial change based on the sandboxers' feedback. They suggested examining the extent and reasons for modifications in the task instead of inquiring potential future actions. We discussed this recommendation within our research team and reached a consensus that this change would enhance the clarity of our study. Concurring with this feedback, we replaced ``would'' with ``did'' in Step 6, allowing us to better comprehend participants' actions within each task's context. 

% \subsection{Data Collection}
% \label{subsec:collection-analysis}

\vspace{-2.5mm}
\subsection{Lab Study}
%%\boldification{ After our study design was concrete, we carried out our lab study.}

We conducted the lab studies over a period of three weeks. 

%%\boldification{ We recruited …….}
\textit{Recruitment:} We recruited undergraduate CS students who were at least 18 years of age and enrolled in software engineering courses at the university. We visited the classes in person and briefed them about the study. We asked those participants interested in the study to answer a survey about their demographic information (age, gender, academic level), resources they used for software engineering (GitHub, ChatGPT), and their experience in programming and software engineering. A total of 41 people responded to the survey.

%%\boldification{ Our final participants were …… }
\textit{Participants:} After filtering out the incomplete responses, we invited 39 people, asking about their availability. We received 30 responses, but only 24 participants showed up for the study. We later discarded the data from 2 participants as their files were corrupted due to a fault in the machines. Out of the final 22 participants, 13 self-identified as men and 9 as women. We randomly assigned each participant to one of the two groups while allowing an even distribution of demographics and number of participants in each group. Participant IDs were assigned with the format `PT-X' or `PC-X' for the Experimental and Control groups, respectively. The participants' demographics are summarized in Table~\ref{tab:demo}. As a token of appreciation, students received a \$20 Amazon gift card.

\begin{table}[!ht]
\footnotesize
    \vspace{-3mm}
    \caption{\label{tab:demo} \small{Participant demographics.}}
    \vspace{-2mm}
    % \centering
    \begin{tabular}{p{4cm}|r|r|r}
    \hline
         & \textbf{Experimental} & \textbf{Control} & \textbf{Total} \\ \hline
        Man & 6 & 7 & 13 \\ 
        Woman & 5 & 4 & 9 \\ \hline
        Sophomore &2&1&3\\
        Junior &5&5&10\\
        Senior &4&5&9\\ \hline
        Enrolled in SE Courses & 3 & 4 & 7 \\ 
        Enrolled in SE Courses + Worked on SE Projects & 8 & 7 & 15 \\ \hline
        Total & 11 & 11 & 22 \\ \hline
    \end{tabular}
    \vspace{-10px}
\end{table}

%%\boldification{ With our participants, we followed this study protocol…}
\textit{Study Protocol:} All studies were in lab sessions at the University lab with up to 3 participants at a time, following the university IRB protocol. The study proceeded as follows: the participants agreed to an IRB-approved informed consent and were briefed about the different stages of the study, then filled out the pre-study questionnaire, performed the three tasks in the study (the Experimental group was also asked to fill out the AAR/AI questions before and after each task), and finally filled out the post-study questionnaire. The sessions were recorded with participants' consent and lasted around 80 minutes each. Before and after each study session, the browser histories and git branches were deleted to prevent unwarranted advantages and ensure all participants could start the session with the same information. %All tasks built on each other, all participants performed the tasks in the same order irrespective of their group.
%All the participants performed the same tasks in the same order, irrespective of their group. 

%% file: table/Metrics-Instruments.tex
\setlength\aboverulesep{0pt}
\setlength\belowrulesep{0pt}
\setlength{\tabcolsep}{2pt}
\newcommand{\PreserveBackslash}[1]{\let\temp=\\#1\let\\=\temp}
\newcolumntype{C}[1]{>{\PreserveBackslash\centering}p{#1}}
\newcolumntype{L}[1]{>{\raggedright\let\newline\\\arraybackslash\hspace{0pt}}m{#1}}
\newcommand{\pb}[1]{\parbox[t][][t]{1.0\linewidth}{#1} \vspace{-2pt}}
\newcommand{\twothreerule}[0]{\cmidrule(l{\tabcolsep}){2-3}}
\footnotesize
\begin{tabular}{L{12.5mm}|L{22mm}L{48mm}}
\toprule
 \textbf{Evaluation} & \textbf{Construct} & \textbf{Metrics, instruments, procedures, and literature} \\ \midrule
\multirow{3}{*}[-0.8mm]{\pb{Students' perceptions of ChatGPT}} 
 % & Usage (RQ1) & Interactions (Frequency of Usage)~\cite{xu2022ide} \\ \twothreerule
 & Cognitive Load (RQ1) & NASA TLX \cite{hart1988development, brachten2020ability,  schmidhuber2021cognitive} \\ \twothreerule
  & Perceived faults (RQ2) & AAR/AI \cite{dodge-aar,khanna2022finding}, Human-AI interaction guidelines \cite{amershi2019guidelines} \\ \twothreerule
 & Continuance Intention (RQ1) & Post-study questionnaire  \\ \midrule
\multirow{3}{*}[-0.8mm]{\pb{Effects of interacting with ChatGPT}} 
 
 & Productivity (RQ1) & Task performance - Correctness~\cite{xu2022ide}, Time to complete~\cite{xu2022ide} \\ \twothreerule
 & Consequences of AI faults (RQ2) &  AAR/AI \cite{dodge-aar,khanna2022finding}\\ \twothreerule
 & Self-efficacy (RQ1) & Self-efficacy questionnaire~\cite{steinmacher2016overcoming} \\ \midrule

\end{tabular}

%% file: table/AARAI-table.tex
\begin{tabularx}{\textwidth}{p{5.3cm}X}
    \toprule
    \textbf{AAR/AI Steps} & \textbf{AAR/AI in our Empirical context} \\
    \midrule
    1. Defining the rules: How are we going to do this evaluation? What are the details regarding the situation? & We briefed the participants about the study details and how we were going to do the evaluation. Then we stated: ``You will be given a questionnaire before and after each task. Please be detailed in your responses as that will help us evaluate ChatGPT's performance.'' 
    \\
    \addlinespace
    2. Explaining the objectives of the AI agent: What is the AI’s objective for this situation? & We oriented the participants about the primary objective of ChatGPT by stating, ``The primary objective of ChatGPT will be to assist you by providing contextual, disambiguous, and correct information.'' \\
    \midrule
    \multicolumn{2}{c}{\textbf{Inner Loop}} \\
    \midrule
    3. Reviewing what was supposed to happen: What did the evaluator intend to happen? & We asked ``What do you think should happen when you use ChatGPT for this task?'' The participants chose between: It will (provide (all/some))/(not provide any) useful information I need to complete the task. \\
    \addlinespace
    4. Identify what happened: What actually happened? & The participants did a task, then we asked ``What actually happened when you used ChatGPT for this task?'' The participants chose between: It (provided (all/some))/(did not provide any) useful information I need to complete the task. \\
    \addlinespace
    5. Examine why it happened: Why did things happen the way they did? & We asked ``Why do you think ChatGPT behaved this way?'' \\
    \addlinespace
    6. Formalize learning (end inner loop): What changes would you make in the decisions made by the AI to improve it? & We asked two questions: ``To what extent did you modify ChatGPT’s responses for solving the task?'' The participants chose between: Did not modify at all/Modified (slightly/significantly). Then, we asked them to ``Briefly explain why?'' \\
    \midrule
    % \textbf{End Inner Loop} \\
    \multicolumn{2}{c}{\textbf{End Inner Loop}} \\
    \midrule
    7. Formalize learning: What went well, what did not go well, what could be done differently next time? & We asked three questions: “What went well?”, “What did not go well?”, “What could be done differently next time?” \\
    \bottomrule
  \end{tabularx}

%% file: sec/3.Results.tex
\vspace{2.5mm}
\section{Results}
\label{sec:RQ1}
%NEED TO ADD A PARAGRAPH HERE
\revised{In the following, we present our findings of participant experiences with and without ChatGPT in the context of performing SE tasks}.

\vspace{-2mm}
\subsection{RQ1: Effectiveness}
\label{subsec:RQ1}
%\textit{How effective is convo-genAI in helping students in software engineering? }

%%\boldification{To address RQ1, we used statistical analysis. We first discuss H1 and H2}
To address RQ1 \revised{(effectiveness of students using ChatGPT for SE tasks)}, we tested our hypotheses (Section \ref{subsubsec: instruments}) using the Mann-Whitney U test~\cite{mcknight2010mann}.\footnote{We performed Shapiro-Wilk's test \cite{shapiro1965analysis} for normality and Levene's test \cite{brown1974robust} for equal variances to determine the suitability of parametric tests. Since not all metrics met the assumptions for parametric tests, we used non-parametric tests for all hypotheses for consistency.} The results are shown on Table \ref{tab:Stats}.

%some metrics met the assumptions for parametric tests, while others did not (refer to the supplemental material \cite{supplemental2023}). To ensure consistency in our statistical reporting, non-parametric tests were used for all hypotheses}. %On analysis, it is found that \textbf{\textit{H1 and H2 are not supported}} (Table \ref{tab:Stats}). 
%in relation to the corresponding null hypotheses, which assume no group differences. 
%We used 
%On the contrary, it was found that the participants using ChatGPT perceived a higher cognitive load (H1), especially in terms of frustration. Moreover, no statistically significant difference in overall productivity was observed (H2).

% ===================================
% Stat Table
% ===================================

\begin{table*}[bhtp]
    % \footnotesize
    % \vspace{-2mm}
%    \begin{threeparttable}
    \caption{Statistical results for cognitive load (NASA TLX) and task productivity (correctness).}
    \label{tab:Stats} 
    \vspace{-4mm}
  \input{table/Stat-Table}
  \begin{tablenotes}
  \scriptsize
\item The estimates, p-values, and Cliff's delta (effect size) are with respect to Mann Whitney U-test. The \textbf{highlighted} columns are statistically significant. Negative $\delta$ suggests that the variable tends to have higher values in the Control group. We consider $|\delta| <$ [0, 0.15) to be no effect, $|\delta| \in$ [0.15, 0.33) to be small, $|\delta| \in$ [0.33, 0.47) to be medium, and $|\delta| >$ 0.47 to be large, by convention \cite{romano2006exploring}. 
    \end{tablenotes}
    \vspace{-1mm}
%\end{threeparttable}
\end{table*}

%%\boldification{TLX results indicate higher cognitive load for the experimental group, especially in terms of frustration.}
To assess H1 (Cognitive Load), we examined the answers to TLX questions. 
As shown in Table~\ref{tab:Stats}, \textbf{frustration levels among participants using ChatGPT were significantly higher} than among those from the Control group (U=101, p=$0.008^{***}$), with a large effect size ($\delta$=0.669). Previous studies in Human-Robot Interaction highlight that end-user frustration is often induced by erroneous behavior in automated systems \cite{weidemann2021role}. Our study corroborated similar patterns, highlighting an association between heightened frustration levels and faults in ChatGPT's behavior and responses. Several study participants clearly illustrated this. PT-3 conveyed, \textit{``[ChatGPT] was fighting me a lot about the whole NOAA thing [Task 1, test case 2].''} Similar challenges were echoed by PT-7, who mentioned that \textit{``[ChatGPT] misinterpreted my questions, was REALLY slow, and didn't account for errors.''} Meanwhile, PT-1 articulated mistrust, \textit{``I could not rely on [ChatGPT] to tell me when functions exist or not''}. Although the other factors were not statistically significant (and had small or negligible effect sizes), the participants using ChatGPT reported a slightly higher Mental demand (Med=15) compared to the Control group (Med=14) and perceived lower levels of Performance (Med=9) compared to others (Med=12). Still, Physical demand was rated very low for both groups, and there was no difference in Temporal nor Effort dimensions between the groups. Overall, \textit{H1 is not supported}: the Experimental group had no significant advantages over the Control group across the cognitive load dimensions. 
% with no observable disparity in perceived Effort (Med=14) across both cohorts. 

%%\boldification{We found no difference in overall productivity although there is practical significance for task 1 and task 2}
With respect to H2 (productivity), we could not find statistical differences in overall productivity (in terms of task correctness) between the two groups (Table \ref{tab:Stats}). We observed a medium effect size pointing that participants using ChatGPT had lower productivity in terms of fixing code functionalities (Task 1: $\delta$=-0.339) and higher productivity in terms of removing code smells (Task 2: $\delta$=0.421). %From these findings, it can be concluded that \textbf{\textit{H2 is partially supported.}} 
From these findings, it can be noted that although there were some variations in task-specific productivity, \textbf{we could not find an impact on productivity for the participants using ChatGPT} \textit{(H2 is not supported)}.
%ChatGPT did impact task-specific productivity to some extent, but did not have any overall impact.

%%\boldification{We found no difference in overall perceived self-efficacy but task-specific trends were observed similar to H2.}
To assess how the allotted resources influenced participants' self-efficacy (H3), we analyzed the pre- and post-study response variations. The Wilcoxon-signed rank test did not show a statistically significant difference comparing the total self-efficacy score before and after ($p(Experimental)=0.214$, $p(Control)=0.7$). 
% However, it is worth noting that the total self-efficacy of 7 out of 11 participants who used ChatGPT increased, while 3 decreased and 1 stayed constant. Most of the participants finished more confident than when they started the study. For the participants who used alternate resources (Control group), an opposite behavior was observed: the score of 6 out of 11 participants decreased after the study, whereas 5 increased.

\begin{figure}[hbt]
\centering
\vspace{-6px}
\includegraphics[width=0.95\linewidth]{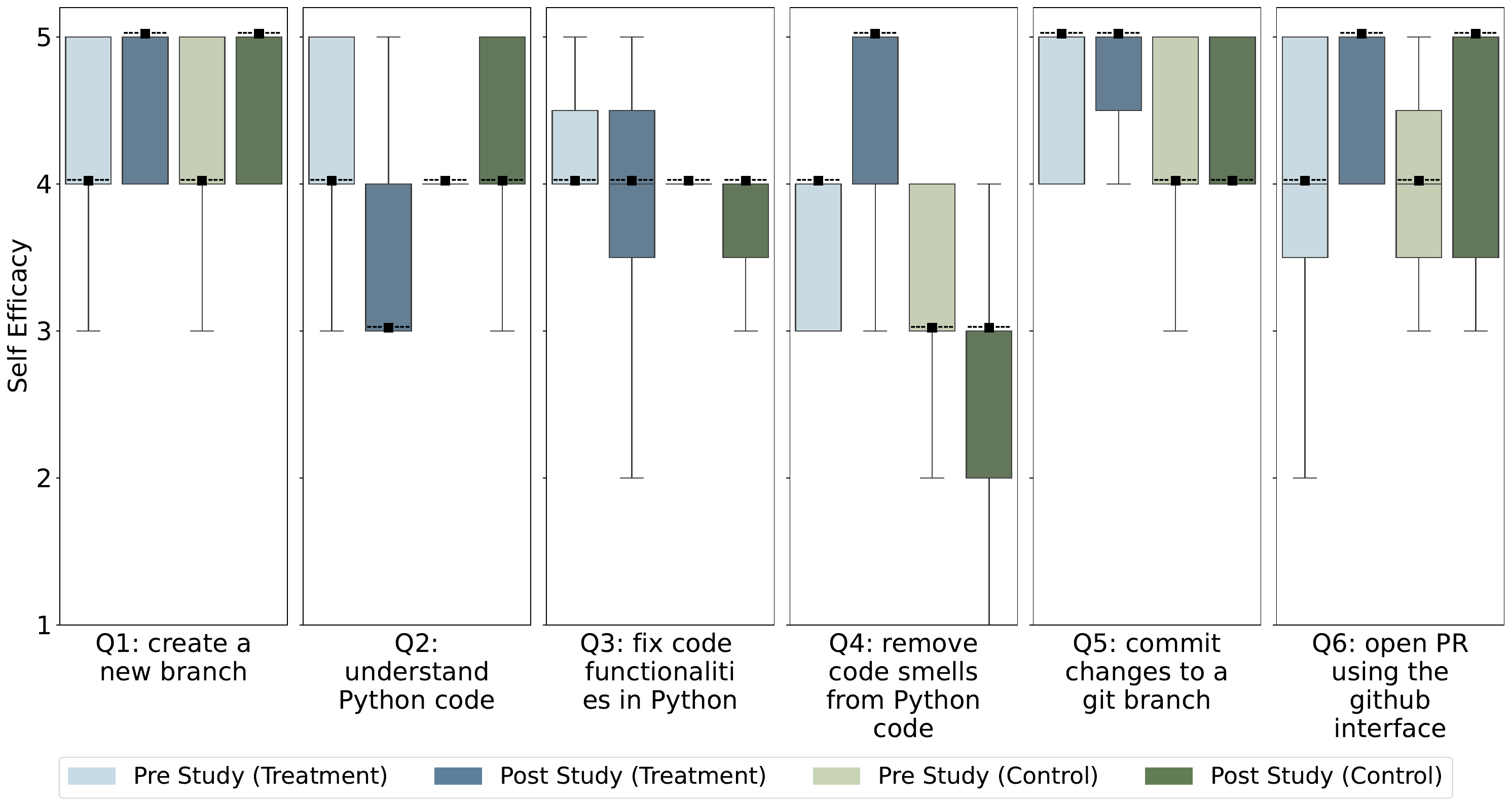}
\vspace{-10px}
\caption{\small{Self-efficacy results (box plots) per question. Medians are highlighted using black dots.} }
\vspace{-10px}
\label{fig:se-results}
\end{figure}

Figure~\ref{fig:se-results} shows the distribution of self-efficacy scores per question. We only noticed differences in terms of distributions and median shifts (comparing before/after for the groups) for Q2 and Q4. Overall, we observed a decrease in the Q2 scores (related to understanding Python code) for those who used ChatGPT, while the distribution remained the same for those who did not use it. For Q4 (related to the removal of code smells), the score for the group using ChatGPT increased after the task, while the median for the Control group remained the same (with a noticeable shift of the distribution towards negative scores). These trends align with the task-specific productivity findings (H2) observed with ChatGPT. For the other items, there were no differences when comparing the groups. We observed similar distributions and median shifts for the git/GitHub items (Q1, Q5, Q6) as well as for Q3 (fixing code functionalities)---comparing before and after for Experimental and Control groups. In summary, although ChatGPT influenced students' self-efficacy for certain items, overall, it \textbf{did not positively influence students' self-efficacy} ({\textit{H3 is not supported}}). 

%We also notice an increasing trend for the question  in the experimental group indicating that the perceived ability to remove code smells increased after using ChatGPT.

%\rudy{shift in the median is identical for both groups in Q1, Q6)} 

%%he median of the before and after answers for questions related to git/GitHub usage (Q1, Q5, Q6) were the same for both groups. 

%No noticeable difference was observed for Q3 (fixing code functionalities). 
%The scores for the question related to understanding Python code decreased for the participants who used ChatGPT and remained the same for those who did not. The decreasing trend indicates that their perceived ability to comprehend Python code decreased after using ChatGPT for the tasks. We also notice an increasing trend for the question related to the removal of code smells in the experimental group indicating that the perceived ability to remove code smells increased after using ChatGPT. These trends align with the task-specific productivity findings \textbf{(H2)} observed with ChatGPT. 
% We concluded that \textbf{\textit{H3 is partially supported}}. 

%%\boldification{Next, we notice mixed continuance intention}

\begin{figure}[hbt]
\centering
\vspace{-10px}
\includegraphics[width=0.9\linewidth]{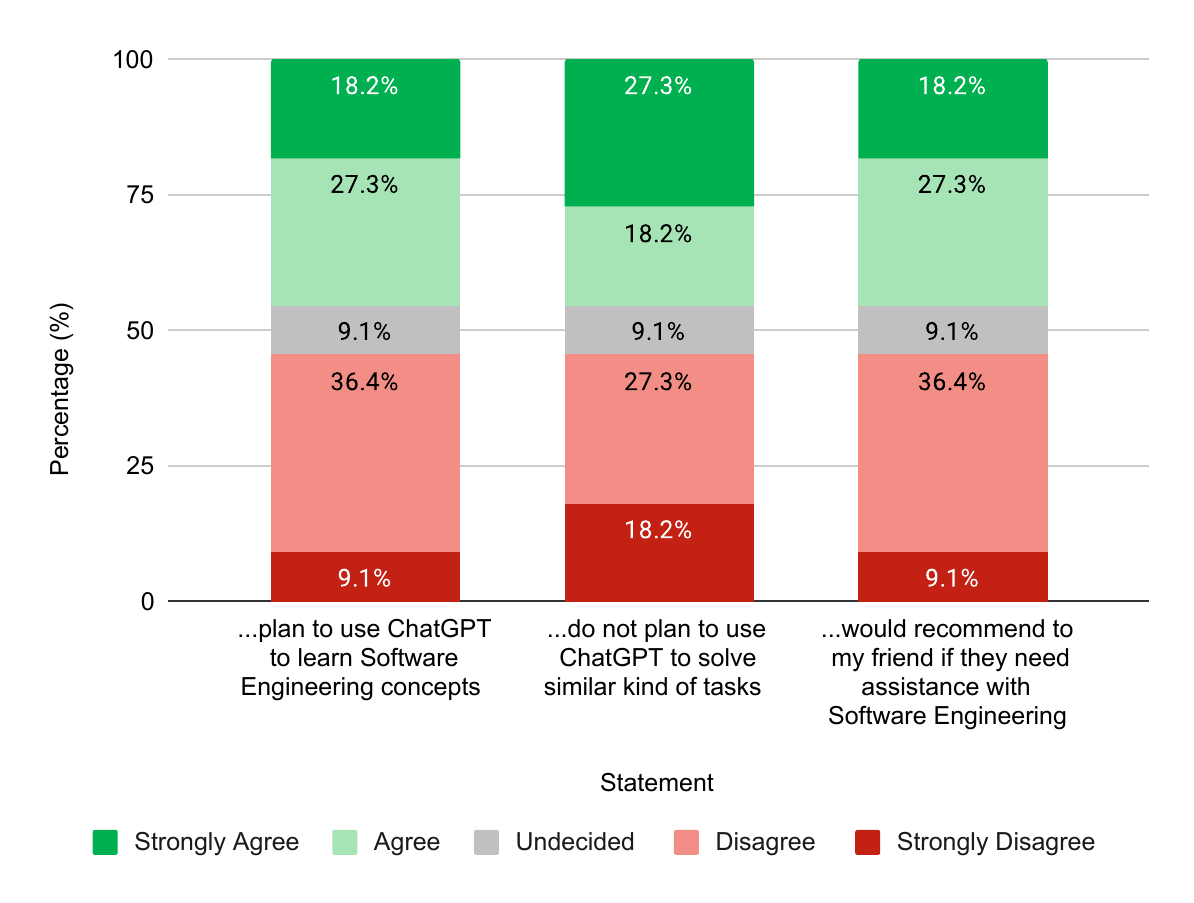}
\vspace{-15px}
\caption{\small {Continuance intention towards using ChatGPT (\%)} }
\vspace{-10px}
\label{fig:intention-results}
\end{figure}

Furthermore, participants' continued intention to use ChatGPT (part of the post-study questionnaire---Table~\ref{table:metrics}) received mixed responses (Figure \ref{fig:intention-results}). One respondent was undecided throughout (9.1\%). The remaining participants were equally divided in their opinion about using ChatGPT and recommending it to friends needing assistance with software engineering. Five participants (45.5\%) showed positive responses, and the other 5 (45.5\%) provided negative responses, indicating a polarized view among students. Despite a segment expressing readiness to leverage the tool, an equal fraction expressed notable resistance: \textit{``I would have liked to be able to ask someone knowledgeable in Python about [task 1] (PT-11)''}.

%2 respondents (18.2\%) strongly agreed, while 3 agreed (27.3\%), making a combined positive response of 45.5\%. On the other hand, 1 respondent (9.1\%) strongly disagreed, and 4 respondents (36.4\%) disagreed, also making a combined negative response of 45.5\%. As for using ChatGPT for similar kinds of tasks, 5 out of 11 respondents (45.5\%) show a positive inclination and an equal percentage (45.5\%) of respondents disagree or strongly disagree, revealing a polarized view. 
%This dichotomy implies a divide among students regarding the potential use of ChatGPT in software engineering. Despite a segment expressing readiness to leverage the tool, an equal fraction expressed notable resistance: \textit{``I would have liked to be able to ask someone knowledgeable in Python about [task 1] (PT-11)''}.

% \noindent\fbox{%
% \parbox{0.97\linewidth}{
% \textbf{O/P:} Higher cog load. Had Boosted task-specific productivity and self efficacy, though no overall impact. Students express mixed intentions regarding continuance of the tool.
% }
% }
\vspace{-2mm}
\subsection{RQ2: Pitfalls}
%\textit{What are the current pitfalls in convo-genAI?}

This section presents the results for RQ2, i.e., the faults made by ChatGPT, their causes, and consequences \revised{within the context of assisting students in SE}. Three authors qualitatively analyzed (open coding) the AAR/AI responses and identified \revised{ChatGPT's} faults and their consequences on the \revised{participants}. The coding was done collaboratively, with the authors engaging in iterative discussions (\revised{over 2 weeks}) to reach a consensus on the final codes. 

\subsubsection{Faults made by \revised{ChatGPT}} 
The faults made by \revised{ChatGPT in the context of assisting students in SE} were grouped into 5 categories: (F1) Limited advice on niche topics, (F2) Inability to comprehend the problem, (F3) Incomplete assistance, (F4) Hallucination, and (F5) Wrong guidance.
% (Table \ref{tab:faults}). 

\begin{comment}
\begin{table*}
    \footnotesize
    \vspace{-2mm}
    \caption{\label{tab:faults} \small{Faults made by convo-genAI (ChatGPT)}}
     \vspace{-8pt}
  \input{table/Faults}
   \vspace{-12pt}
\end{table*}
\end{comment}

%%\boldification{These are the faults and we now find the evidence behind RQ1 results}
\textbf{F1: Limited advice on niche topics} [PT-1, 5, 9]. ChatGPT struggled to provide expert advice on topics specific to a niche (e.g., a domain, a library, or a concept). For instance, PT-1 explains: \textit{``ChatGPT only was helpful with general info, like git commands or logic issues. It wasn't helpful with niche specifics, like discerning between functions to use in a Python library''}. According to our participants,\textit{ ``for anything that wasn't super standard, ChatGPT struggled to easily give useful answers. (PT-1)''} and thus \textit{``having it define or explain ambiguous concepts did not help much (PT-5)''}. Moreover, ChatGPT provided limited advice regarding Python code functionalities (PT-1, 5, 9). This could be a reason for the \textbf{observed decrease in Task-1 specific productivity} for participants using ChatGPT \revised{for the study tasks}. PT-9 said, \textit{``ChatGPT did not have as much knowledge about the NOAA python library, and confidently told me incorrect ways to `fix' my code.''}

%%Igor: F2: ChatGPT did not understand the thing % rudy: Added a line to differentiate.

\textbf{F2: Inability to comprehend the problem} [PT-1, 5-8, 10]. ChatGPT could not always understand the participants' goals and the problems they were facing. Participants revealed that \textit{``it incorrectly identified nonproblems as problems and missed actual problems (PT-1)''} and did not \textit{``know the exact thing you want it to do despite giving it context (PT-6)''}. Before starting Task-1, 6 out of 11 participants (54\%) responded that ChatGPT would provide all the required information. However, after completing the task, all participants marked that it only provided some information. This mismatch in expectations significantly \textbf{increased the participants' frustration levels}. PT-7 emphasized,\textit{ ``[ChatGPT] misinterpreted my questions at times, was REALLY slow, and did not account for errors in code it provided me''.} 

%%%Igor: Sometimes when F2 happen, the assistance was incomplete. % rudy - Done

\textbf{F3: Incomplete assistance} [PT-1-3, 6-9, 11]. ChatGPT sometimes provided incomplete and partially correct assistance even when it was able to grasp the problem. PT-11 pointed out, \textit{``I did ask ChatGPT questions about completing the task, but it did not give me answers on how to solve the whole task''}. Additionally, participants discovered that \textit{``[ChatGPT] knows some things and can help give you advice on those things, but it won't immediately give you the correct answer (PT-6)''}. They also pointed out that, \textit{``Some code provided by ChatGPT was correct, while some were incorrect and required modifying (PT-9)''}. This AI behavior likely \textbf{affected participants' mental workload}. As noted by PT-11, \textit{``I could not figure out how to fix the import problem, and ChatGPT's suggestions didn't work''.}

%%% Make F4 a$
% \textbf{F4: Cannot Access Relevant Information}: ChatGPT did not have access to relevant information needed for assisting the users. PT-4 commented that \textit{``it [ChatGPT] did not have access to the documentation for the packages we were using''}. Similarly, PT-6 indicated that, \textit{``It doesn't fully understand the documentation/interactions of some libraries/functions.''}

\textbf{F4: Hallucination} [PT-4, 9, 11]. ChatGPT tends to hallucinate, creating false answers when it does not know the correct solution. Participants pointed out multiple instances of this. 
PT-4 stated that when ChatGPT \textit{``did not have access to the documentation for the packages\ldots it hallucinated answers''} and that it \textit{``made up parameters for functions that were unfamiliar''}.
Similarly, PT-11 noted that \textit{``it did hallucinate sometimes, said there was a way to use a function in the noaa-sdk that was not possible''}. Additionally, there were instances of \textit{confirmation bias} (when the AI conforms to the users' statements/requests, regardless of the actual accuracy/feasibility). PT-4 highlighted that ChatGPT \textit{``was biased towards a `yes, there are code smells' response...When they don't exist, it hallucinates them.''}

\textbf{F5: Wrong guidance} [PT-2-4, 7-11]. In addition to hallucinating, there were other instances where ChatGPT gave wrong guidance, or \textit{``incorrect ways to fix [problems] (PT-9)''}. For example, when it could not comprehend the problem (F2) participant (PT-8) was facing, it gave a piece of incorrect advice: \textit{``It couldn't figure out test case 3 and kept telling me to check my drivers...without realizing there were missing imports (PT-8)''}. ChatGPT often %\textit{``made mistakes (PT-2)''} and 
\textit{``did not account for errors in [solutions] it provided (PT-7)''}. It also suggested \textit{``incorrect ways to fix [problems] (PT-9)''} when it had limited knowledge on a topic (F1) and hence some of its solutions appeared \textit{``evidently wrong or unnecessary (PT-10)''}.
 
%... PT-10 stated that ChatGPT \textit{“clearly gave wrong guidance”} and some of its responses looked \textit{“evidently wrong or unnecessary”}. Furthermore, ChatGPT often misguided the users when it was uncertain or struggled to discern the real issue: \textit{``ChatGPT did not have as much knowledge about the NOAA python library, and confidently told me incorrect ways to ``fix`` my code. (PT-9)``, 

In summary, we observed that Experimental group participants' mental load and frustration increased when ChatGPT was unable to comprehend the problem (F2) or provided incomplete assistance (F3). Additionally, they perceived equal effort as the Control group participants, frequently tasked with identifying and resolving errors when ChatGPT hallucinated (F4) or provided wrong guidance (F5). %\rudy{As described in F2, F3}

%\boldification{On using HAI guidelines as a lens, we found that 5 guidelines were reported to be in violation.}
\subsubsection{Causes of these faults and their consequences} 
\begin{figure*}[!bht]
\centering
% \vspace{-10px}
\includegraphics[width=0.9\textwidth]{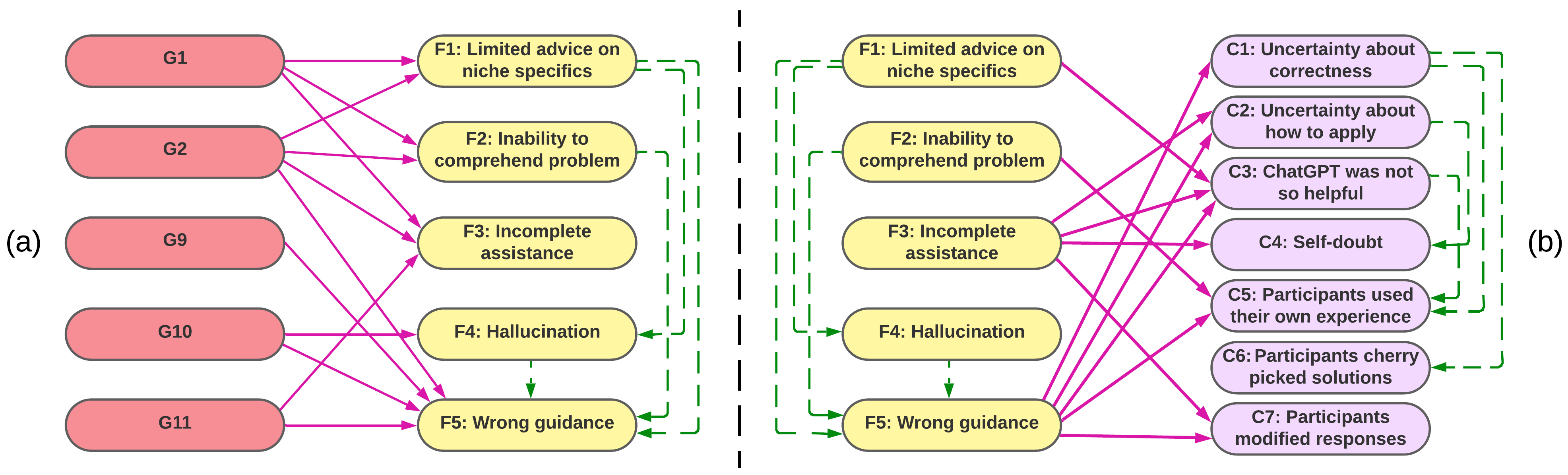}
\vspace{-10px}
\caption{\small {The (a) causes and (b) consequences of ChatGPT's faults: Violation of Human-AI Interaction guidelines (G1, G2, ...) led to faults (F1, F2, ...). Faults had a cascading effect: one led to another and further led to consequences (C1, C2, ...) for participants. Some of these consequences led to other consequences.} }
\vspace{-5px}
\label{fig:cause}
\vspace{-8px}
\end{figure*}

For each of these faults, we examined why they occurred and the consequences they had on the participants (Figure~\ref{fig:cause}). 
As mentioned in Section~\ref{subsubsec: instruments}, \revised{participants rated Human-AI (HAI) guideline statements specific to ChatGPT interactions at the end of the experiment, which was used to assess guideline violations.}
% As mentioned in Section~\ref{subsubsec: instruments}, participants were asked to identify which Human-AI (HAI) guidelines ChatGPT violated at the end of the experiment. 
% \revised{We manually validated all reported numbers with the open-ended text responses, and found no discrepancy (the text matched their reported numbers)}.
\revised{We manually triangulated the response scores with open-ended text responses and found no discrepancies.}

\begin{figure}[!hbt]
\centering
\vspace{-8px}
\includegraphics[width=0.85\linewidth]{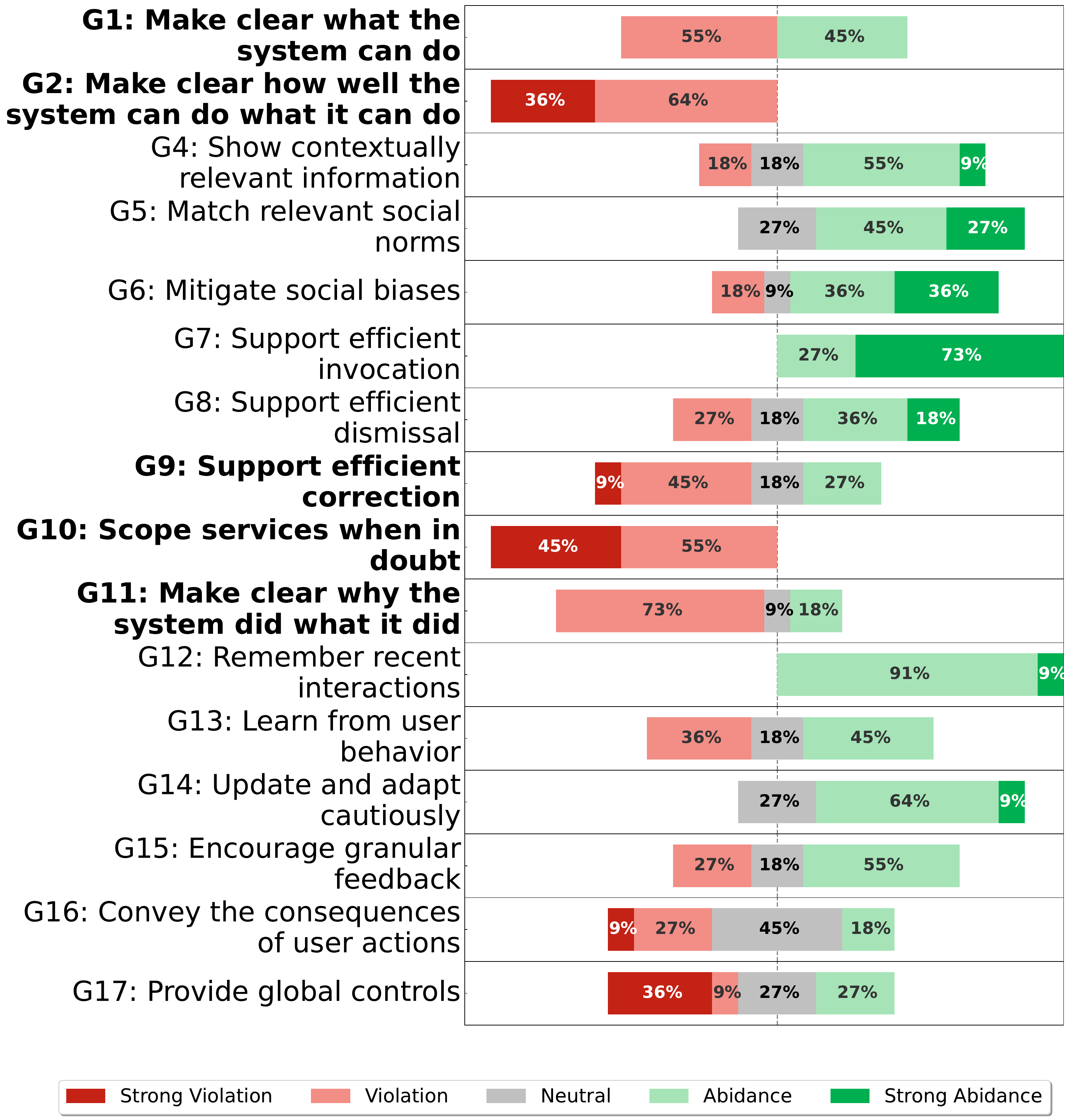}
\vspace{-10px}
\caption{\small {Human-AI Interaction guideline violations reported by participants; those found by more than 50\% are in bold.}}
\vspace{-9px}
\label{fig:guideline}
\end{figure}
% For each of these faults, we examined why they occurred and and what consequence it had on the students. (Figure \ref{fig:cause}). Recall, participants mentioned which of the \textbf{Human AI guidlines} ChatGPT violated at the end of the experiment (see Section \ref{subsubsec: instruments}). 

\begin{comment}
\begin{table*}[bth]
    \footnotesize
    % \vspace{-2mm}
    \caption{\label{tab:consequences} \small{Consequences arising from convo-genAI's (ChatGPT) faults}}
     \vspace{-8pt}
  \input{table/Consequences.tex}
   \vspace{-12pt}
\end{table*}
\end{comment}

\textbf{Causes:} All participants identified violations of Guideline 2 (G2: Make clear how well the system can do what it can do) and Guideline 10 (G10: Scope services when in doubt). Additionally, 8 out of 11 participants ($\sim$73\%) found violations in Guideline 11 (G11: Make clear why the system did what it did) [PT-1-3, 5-7, 10-11], and 6 out of 11 participants ($\sim$54.5\%) found violations in Guideline 1 (G1: Make clear what the system can do) [PT-1, 2, 4-6, 11] and Guideline 9 (G9: Support efficient correction) [PT-2-4, 7, 8, 11] (Figure~\ref{fig:guideline}). For each participant, we then mapped the faults (reported in AAR/AI) to the guideline violations they reported. 

%\boldification{These violations answer why faults occurred...}

 ChatGPT was perceived as offering limited advice on niche topics (F1), being unable to comprehend problems (F2), and providing incomplete assistance (F3), as no appropriate expectation of quality was set. Microsoft's HAI Guidelines 1 and 2 focus on clarifying expectations to prevent mismatches between users and AI, as demonstrated in prior literature \cite{li2022assessing}. Thus, it is likely that these faults resulted from ChatGPT's initial shortcomings in clearly stating its capabilities (G1 violation) and inadequately indicating how often it might make mistakes in its responses (G2 violation). There were instances when ChatGPT did not support efficient correction (G9 violation), making it difficult to refine its responses when it was incorrect: \textit{“It was fighting me a lot...(PT-3)”}. Furthermore, ChatGPT also provided ambiguous/wrong information without conveying its uncertainty (G10 violation) and made it hard to gain an explanation regarding its decision-making process (G11 violation), likely resulting in hallucination (F4) and wrong guidance (F5). 
% Comment for Anita: It did not scope neither did it explain how it got to a solution, that resulted in wrong guidance.

%\boldification{Moreover, \textbf{faults had a cascading effect} - it caused more faults.}
\textbf{Cascading faults:} We also found that these faults had a cascading effect, where one fault led to another (green arrows in Figure~\ref{fig:cause}). For instance, when ChatGPT struggled with niche specifics (F1) or was unable to comprehend problem (F2), it hallucinated (F4) and provided wrong guidance (F5): \textit{``ChatGPT did not have as much knowledge \ldots and confidently told me incorrect ways to `fix` my code. (PT-9),'' ``kept telling me to check my drivers\ldots without it realizing there were missing imports. (PT-8).''}

%\boldification{We identified 7 consequences. Here we cover what they mean, what causes them, and how effects productivity, cog.load, and self-efficacy. (Linked with RQ1)}

\textbf{Consequences:} From the AAR/AI responses, we identified 7 consequences for participants that arise due to ChatGPT's faults and grouped them into 3 categories: \textit{Uncertainty} (uncertainty about correctness, uncertainty about how to apply), \textit{Reflections} (ChatGPT was not so helpful, self-doubt), \textit{Actions} (\revised{participants} used their own experience, cherry-picked solutions, and modified the responses).% (see Table \ref{tab:consequences}) :

%When participants were \textit{unsure}, their perceived \textbf{mental workload increased}, as they grappled with how to apply the responses and questioned their correctness. 
Participants grappled with how to apply \revised{ChatGPT's} responses because of \textit{uncertainty} and questioned its correctness:

%There were instances of \textit{uncertainty} (C1, C2) when participants grappled with how to apply \revised{ChatGPT's} responses, questioning their correctness:

\textbf{C1: Uncertainty about correctness} [PT-1, 2, 6, 10]. Participants were uncertain about the correctness of the responses provided by ChatGPT. For example, PT-1 stated, \textit{``I could not rely on it to tell me when functions exist or not''}. PT-2 expressed skepticism, stating, \textit{``I don't think it provided fully correct data, so I am inclined to pick parts only''}. This was related to the wrong guidance (F5) provided by ChatGPT, as per PT-10: \textit{``Some answers look[ed] evidently wrong or unnecessary. It is important to modify the code based on my experience.''} 

\textbf{C2: Uncertainty about how to apply} [PT-1-3]. Participants were confused about how to apply/use the information provided by ChatGPT. As PT-1 expressed, \textit{``I didn't use ChatGPT's responses since I wasn't sure how to apply them''}. We noticed that this consequence stemmed from two faults: incomplete assistance (F3)---\textit{``got confused over suggestions w/ the weather library…correction that wasn't made obvious to me for palindrome (PT-2)''---}and wrong guidance (F5)---\textit{``I'm not super sure why this didn't work. It was fighting me a lot about the whole NOAA thing (PT-3).''}

%Participants found ChatGPT not so helpful and occasionally doubted themselves. With these \textit{reflections}, participants \textbf{perceived lower levels of performance}, and their \textbf{self-efficacy} related to Python code comprehension \textbf{decreased} after using ChatGPT.

% Other than these consequences, participants provided \textit{reflections} (C3, C4) about why ChatGPT was not so helpful, making them occasionally doubt themselves. 
% With these \textit{reflections}, participants \textbf{perceived lower levels of performance}, and their \textbf{self-efficacy} related to Python code comprehension \textbf{decreased} after using ChatGPT.

Participants' \textit{reflection} (C3, C4) revealed that they found ChatGPT not so helpful and occasionally doubted themselves:

\textbf{C3: ChatGPT was not so helpful} [PT-1, 5, 10, 11]. We found instances where participants thought that ChatGPT was not so helpful. This was because ChatGPT provided limited advice on niche topics (F1), with PT-5 noting \textit{``having [ChatGPT] define or explain ambiguous concepts did not help much''}. Other reasons included it providing incomplete assistance (F3), \textit{``\ldots ChatGPT's suggestions didn't work (PT-11)''} and wrong guidance (F5): \textit{``The suggestions...are not so helpful. It clearly gave wrong guidance (PT-10)''}.

\textbf{C4: Self-doubt} [PT-2, 4]. Participants doubted themselves, suspecting that they might be at fault. For instance, PT-2 shared, \textit{``I got confused over suggestions with the weather library, likely I should have provided the full error... And I also may have asked something wrong? correction wasn't made obvious to me (PT-2)''}. This surfaced when ChatGPT delivered incomplete assistance (F3), leaving participants \textbf{uncertain about how to apply} the suggested solutions. 

 Participants reported that they had to undertake \textit{actions} (C5, C6, C7) to tackle portions of tasks on their own \revised{(no reliance)}, cherry-pick from provided solutions and modify responses provided by ChatGPT \revised{(partial reliance)}:
%These \textit{actions} added to the participants' \textbf{mental workload}, and explains why overall task \textbf{ productivity was not positively influenced}.

\textbf{C5: \revised{Participants} used their own experience} [PT-1, 4, 7, 10]. Participants had to use their own experience to tackle certain aspects of the tasks, particularly when ChatGPT was unable to comprehend their problem (F2) and thus was perceived to be \textbf{not so helpful}: \textit{``Everything else I had to do on my own because ChatGPT didn't comprehend enough to help (PT-1)''}. Participants devised their own solutions when they were \textbf{uncertain of the correctness} of ChatGPT’s suggestions, attributed to its tendency to provide wrong guidance (F5): 
\textit{``Some answers look[ed] evidently wrong or unnecessary. It is important to modify the code based on my experience (PT-10)''}.

\textbf{C6: \revised{Participants} cherry-picked solutions} [PT-2, 3, 5]. Participants picked parts of solutions provided by ChatGPT that seemed correct: \textit{``Some of the things I didn't necessarily agree w/ but some of it was valid, so I picked \& chose what I liked (PT-3)''}. When \textbf{uncertain about correctness}, participants were \textit{``inclined to pick parts only (PT-2)''}.

\textbf{C7: \revised{Participants} modified the responses} [PT-6, 7, 9, 10]. \revised{Participants} modified the responses provided by ChatGPT to come up with the correct solution. PT-7 noted, \textit{“I used its responses in tandem so I kind of combined them. When it wasn't right I did it myself”}. This consequence stemmed from ChatGPT’s incomplete assistance (F3) and also wrong guidance (F5): \textit{``Some code provided by ChatGPT was correct, while some was incorrect and required modifying it (PT-9)''}.

%% file: table/Stat-Table.tex
\definecolor{Cream}{RGB}{254, 238, 214}
% \definecolor{Cream}{gray}{0.85}
% mathbf to bold inside $$
% \footnotesize
\small
\begin{tabular}{l|cccccc|cccc}
    \hline
         & \multicolumn{6}{c|}{NASA TLX} & \multicolumn{4}{c}{Task Correctness} \\ 
         & Mental & Physical & Temporal & Performance & Effort & \cellcolor{Cream} \textbf{Frustration} & Task 1 & Task 2 & Task 3 & Overall \\ \hline
        Estimate & 47 & 51.5 & 64.5 & 45.5 & 45.5 & \cellcolor{Cream} 101 & 40 & 86 & 71.5 & 71 \\ 
        p-value & 0.388 & 0.557 & 0.817 & 0.339 & 0.337 & \cellcolor{Cream} $\mathbf{0.008^{***}}$ & 0.16 & 0.078 & 0.303 & 0.507 \\
        Cliff's delta$(\delta)$ & -0.223 & -0.149 & 0.066 & -0.248 & -0.248 & \cellcolor{Cream} \textbf{0.669} &\textbf{ -0.339} & \textbf{0.421} & 0.182 & 0.174 \\ \hline
        \multicolumn{11}{l}{\textbf{Median values for each group}} \\\hline
        Experimental & 15 & 1 & 15 & 9 & 14 & \cellcolor{Cream}14& 1 & 1 & 2 & 4 \\ 
        Control & 14 & 3 & 15 & 12 & 14 & \cellcolor{Cream}9 & 2 & 0 & 2 & 4 \\ \hline
\end{tabular}

%% file: sec/4.Discussion.tex
\section{Discussion: Recommendation}
\label{sec:discussion}
% Each of these paragraphs have this structure now: study findings & premise -> design recommendations for future using techniques of the past.

% \rudy{Prompt Engineering thoughts to add iff needed: Users should receive clear guidance regarding the effective and appropriate use of these language models. Furthermore, it's crucial to develop tailored strategies that enhance communication with these LLMs, particularly designed to foster educational outcomes.} 

\noindent{\textit{\textbf{It is necessary to customize Generative AI as an effective scaffolding learning agent for software engineering.}}} 
While genAI tools have proven effective in providing quick solutions to user queries, this approach conflicts with the traditional goals of education. Directly giving away answers can diminish the need for critical thinking and impact learning, potentially leading to reduced self-efficacy~\cite{dehghani2011relationship} and motivation~\cite{chang2014effects}. In our study, we observed that participants' self-efficacy decreased in certain cases, like understanding Python code (see Sect. \ref{subsec:RQ1}, Fig. \ref{fig:se-results}). This decline may be attributed to students viewing these tools as advanced search engines that offer ready-made solutions, a practice that instructors fear could impede genuine learning~\cite{lau2023ban}. 

Hence, future research should investigate how genAI can be tailored and optimized as an effective scaffolding learning agent. To be effective, a scaffolding agent must correctly interpret the students' intentions, an aspect where genAI shows promising results. Nonetheless, more work is necessary to understand the expectations of students and instructors and how students express their expectations and engage in dialogue with the agent. These insights can help design agents that grasp students' intentions and adapt their interactions to enhance pedagogical outcomes. Further research can also explore how genAI can be leveraged for personalized student assistance, using techniques like the `persona prompt pattern'~\cite{white2023prompt} to adjust content based on expertise levels. Moreover, future research should consider incorporating pedagogical scaffolds (templates, heuristics, or human intervention) into AI-generated content to clarify the AI's problem-solving process and handle underperforming scenarios. 

% and fostering student-AI collaboration, akin to think-pair-share \cite{kaddoura2013think}. 

\noindent{\textit{\textbf{Recommendations for future Generative AI design}:}} Participants using ChatGPT in our study considered that it violated 5 out of the 18 HAI guidelines. These participants also perceived lower performance levels and uncertainty, which led to self-doubt and possibly lowered their self-efficacy: \textit{``I got confused over suggestions...I may have asked something wrong. (PT-2)'', ``I'm not super sure why this didn't work (PT-3)''}. When participants were \textit{uncertain}, they had to use their own experience (C5) in addition to cherry-picking solutions (C6) and modifying \revised{ChatGPT's} responses (C7) to fit their context. These likely added to the participants' \textit{mental workload} and possibly explain why we could not observe a positive impact on \textit{task productivity}.

Prior literature highlights that implementing Microsoft’s HAI Guidelines has substantiated effects on users, including \textit{increased trust, decreased suspicion} along with them \textit{feeling more in control, less inadequate, more productive, secure, and less uncertain}~\cite{li2022assessing}. Based on this, we suggest that an iterative participatory approach~\cite{simonsen2010iterative} should be followed in the future design of genAI systems to ensure that the systems adhere to the HAI guidelines it currently violates: clearly stating capabilities (G1) and limitations (G2), supporting efficient correction (G9), scoping services when in doubt (G10), and maintaining transparency in the decision-making process (G11). 
Furthermore, \citet{wang2023investigating} recently identified specific design strategies for genAI systems based on their design probe study. They found that communicating AI performance via usage statistics, offering indicators of model mechanisms to support evaluation, and allowing users to configure AI by adjusting preferences helped set proper expectations and inform appropriate usage of AI tools. We speculate that incorporating these design practices and strategies can also mitigate the observed negative consequences on students and significantly foster appropriate levels of trust~\cite{johnson2023make} in genAI-based scaffolding tools.

%\textbf{NEEDS WORDSMITHING and a good ending}

% \Anita{wordsmith: with the advent of AI, there is going to be a big divide between those who can harness AI and those who are not. AI systems like any other tool can embed biases. For example, genderHCI research has found that tools don't support diverse cognitive styles. And when individuals whose style is not support use the AI tool features they face cogntive bias bugs-- an additional cognitive tax that users have to pay. For example...tinkering learning style...}
\noindent{\textit{\textbf{Building Inclusive Technology}:}} 
%With the advent of AI, a new digital divide \cite{cullen2001addressing} may emerge between those who can harness AI and those who cannot. 
Like any other tool, AI systems can embed cognitive biases~\cite{waldman2020cognitive} arising from a lack of support for cognitive diversity~\cite{padala2020gender}. In our study, we found considerable disparities in perceived violations of the HAI guidelines on disaggregating participants’ data based on their gender. All women reported a violation of Guideline 11, perceiving ChatGPT's decision-making process as hard to explain, whereas only 3 out of 6 men reported this violation. This could be because women tend to favor comprehensive information-processing style \cite{meyers2015revisiting} and ChatGPT lacked transparency in its decision-making process. 
Individuals with this information processing style tend to seek out all the information needed before starting a task, whereas those who are selective information processors take the first piece of actionable advice and work on it. 
Similarly, a majority of men marked that ChatGPT did not personalize their experience by learning from their actions over time (4 out of 6 men perceived Guideline 13 violation) and did not allow for global customization of its behavior (5 out of 6 men perceived Guideline 17 violation). In contrast, no women found these guidelines to be in violation. This disparity was likely because of the limited tinkering allowed around its interface and technology. 

Gender HCI research~\cite{beckwith2006gender,santos2023designing} has found that tools often lack support for diverse cognitive styles. As a result, individuals whose styles are not accommodated face \textit{cognitive bias bugs}---an additional cognitive tax they pay when they use the tool. Further, individual differences in how people solve problems and use software cluster by gender~\cite{2016GenderMagInclusiveness}; i.e., some styles are favored more by men than women, and vice-versa. Research spanning a decade has identified that women are often more risk-averse than men~\cite{charness2012strong} and prefer process-oriented learning and are thus less likely to tinker~\cite{beckwith2006tinkering, burnett2010gender}. This implies that if future AI tools are not inclusive of cognitive styles, they will not be inclusive of gender. Previous research has also reported that user interactions and experiences with AI systems are significantly diverse for diverse users~\cite{anderson2021diverse}. Indeed, the current genAI systems have faced criticism for their potential negative impacts on equity~\cite{bender2021dangers, liang2021towards}. In future research, it would be valuable to explore how uncertainty, similar to what we identified in our study, may be associated with cognitive styles. Understanding these connections could pave the way for implementing specific strategies that effectively address and mitigate the impact of uncertainty in the context of AI-assisted learning. To achieve this, the GenderMag method~\cite{2016GenderMagInclusiveness} can be applied for evaluating AI systems throughout their iterative design cycles. Past work has shown that fixing issues found from using GenderMag-based processes creates more inclusive tools and environments~\cite{2016GenderMagInclusiveness, Burnett2016JrnlFinding, padala2020gender,santos2023designing}.

%% file: sec/5.Related-Work.tex
% \vspace{-7px}
\section{Related Work}
\label{sec:background}
% \vspace{-2mm}

% \noindent{\textit{\textbf{{Chatbots in Software Engineering education}:}} 
Chatbots have been popular in educational settings \cite{kuhail2023interacting, wollny2021we, Okonkwo2021chatbots}. In software engineering education, chatbots have been proposed as a way to support students outside of the classroom \cite{binkis2021rule}: offering expert advice in problem-solving activities \cite{verleger2018pilot} and accompanying students in their capstone projects \cite{gonzalez2022improving}. However, these
chatbots are often confined to queries anticipated by educators \cite{farah2022impersonating}, thereby serving as a programmed search feature for frequent queries.

% \noindent{\textit{\textbf{Generative AI in computer science education}: }} 
Generative AI models have demonstrated an impressive ability to solve a large number of computation problems from natural language prompts \cite{chen2021evaluating}, and are now being used for programming tasks \cite{li2022competition, xu2022systematic}. Recent literature highlights how these models outperform most students on typical CS1 and CS2 exam problems \cite{finnie2023my, finnie2022robots}, handle variations in problem-wording \cite{finnie2022robots}, and even surpass human performance on programming competitions \cite{li2022competition}. 
Researchers have further explored how these models can be used to enhance learning computer science. Sarsa et al. \cite{sarsa2022automatic} used genAI to create coding exercises and explanations, both of which can be used to provide practice and guidance to students. Another study combined Codex with learnersourcing (crowdsourcing for learners) to create and validate exercises that are engaging for learners \cite{denny2022robosourcing}. 

Generative models have also been used to aid instructors by automating content creation for interactive course materials \cite{macneil2022generating, macneil2023experiences}. Further, Denny et al. \cite{denny2023conversing} showed that prompt engineering is effective in improving AI responses and thus could give instructors more control to improve the relevancy of generated materials.

Another line of research evaluates the impact of genAI tools on students. 
Prather et al. \cite{prather2023s} studied students' initial impressions of using Copilot for CS1 programming tasks. In a different study, Kazemitabaar et al. \cite{kazemitabaar2023studying} conducted an experiment comparing pre-college students learning Python with and without Codex's help. Their findings indicated that Codex users improved their coding skills more than their counterparts, though both groups achieved similar conceptual comprehension. However, Bird et al. noted that while Codex (Copilot) expedited code writing, it compromised learners' code comprehension \cite{bird2022taking}. Moreover, Vaithilingam et al. \cite{vaithilingam2022expectation} reported that despite the initial user interest in genAI, these tools did not enhance users' task efficiency (time) or accuracy.  
% These observations are consistent with the findings of our study, where participants’ self-efficacy in understanding code dropped after using ChatGPT whereas overall productivity had no improvement. 

Our work complements these, as it looks specifically into supporting students in software engineering.

%% file: sec/6.Threat.tex
% \vspace{-5px}
\section{Threats to Validity}
\label{sec:threat}

%This section discusses the threats to validity relevant to our investigation and the steps we took to mitigate them.

\textbf{\textit{Construct Validity:}}
% instruments from the literature
We used instruments from the literature~\cite{hart1988development,khanna2022finding,amershi2019guidelines,xu2022ide,steinmacher2016overcoming} to measure our constructs as much as possible since they had already been used and validated in other contexts. 
We assessed the instruments adapted to our context with sandbox sessions and refined the instruments and research protocol until the team was confident of the instruments' reliability. 
Nevertheless, despite our efforts, we acknowledge that questions might be misinterpreted and can lead to incorrect measurements.

\textbf{\textit{Internal Validity:}}
We acknowledge that our study, like others, can have self-selection bias, where participants interested in the topic of the study were motivated to participate. 
Participant exhaustion and distraction might have also affected the study results. We mitigated this threat by limiting the length of sessions (80 minutes) and time-boxing the tasks. However, this could mean that participants did not have enough time to complete the tasks. Since participant interaction was the primary focus and time-boxing was applied to both treatments, this does not impact the validity of between-group comparisons. 
Another potential threat is task selection, where study tasks can be too easy or too complicated for our target population. We mitigated this risk by designing the tasks based on the instructor's insights and course materials and sandboxing them with participants of varying expertise. 
\revised{Further, participants assessed the ChatGPT interactions---our investigation focus---to identify faults and guideline violations, which makes the findings dependent on their capabilities. We believe this is not a problem since participants had prior experience with ChatGPT and Python (an average of 3.75 years), and thus had the needed experience to assess their interactions with ChatGPT. Past works \cite{dodge-aar, khanna2022finding} that employed similar instruments have recruited participants with at least 10 hours of experience, which puts us in line with them.}
Finally, desirability bias may have an impact because participants may have favored ChatGPT due to its hype. To mitigate this threat, we adapted neutral and non-judgmental language to frame the questions and explain the experiment and analyzed data cautiously, acknowledging the potential presence of desirability bias when interpreting the results.
%. This means that the participants might have been more motivated or interested in the topic of the study than the general population and thus volunteered to participate. Participant exhaustion and distraction might have also affected the study results. We mitigated this threat by limiting the length of sessions (max of 80 minutes), time-boxing tasks, and reducing redundancy in the questionnaires. We acknowledge that the time allocated to each task might not be enough time to complete the task. However, the constraint was equally applied to both groups, so it does not threaten the validity of between-group comparisons. 
%Another potential threat might be the selection of inadequate tasks or making tasks too hard and complicated for our target population. We mitigated this risk by designing the tasks based on the instructor's insights and course materials and sandboxing them with participants of varying expertise before proceeding to the actual study. 

% Still, ChatGPT (and convo-genAI in general) is a relatively new tool, and participants may have suffered from learning effects, which does not necessarily occur with the Control group. However, we did not observe this effect in our sandboxing sessions and students in general are already using this platform for personal and professional purposes. 

% prior knowledge?

\textbf{\textit{Reliability:}}
%Replicating qualitative results is difficult as human behaviors and perceptions are subject to change over time. 
Interpreting qualitative data can be challenging and potentially affect the study's validity. To ensure consistency, we employed robust techniques from existing literature and continually compared our analysis with established codes. Additionally, we held frequent meetings to discuss and refine the codes and categories until we reached a unanimous agreement.

\textbf{\textit{External Validity:}}
We structured our tasks in Python, which was used in the software engineering courses and students were most familiar with it. Therefore, we trade off generalizability for depth in our context, and our findings might not necessarily generalize to other programming languages, universities, \revised{convo-genAI systems}, and work contexts. \revised{Further, different interaction styles with ChatGPT can lead to different outcomes, and the participants in the study might not have represented all styles.} The relatively small sample size of 22 participants is also a threat to the generalizability of the study. 
We mitigated this threat by recruiting participants from multiple software engineering classes and evenly distributing them in each group based on their demographics. 
%We guided our efforts toward ensuring that the sample was more representative of the wider population and that any demographic differences were minimized.
% Finally, the presented results are related to Generative AI supported Software Engineering scaffolding, and we do not expect that all our findings will apply to other contexts.
% single university, culture

%% file: sec/7.conclusion.tex
% \vspace{-7mm}
\section{Conclusion}
\label{sec:conclusion}
%\boldification{What we found in this study}
Our work comprehensively evaluates convo-genAI's potential and pitfalls in supporting 
software engineering tasks. Our analysis did not reveal any statistical differences in \revised{participants'} productivity or self-efficacy in using \revised{ChatGPT for the study tasks} compared to traditional resources. 
% These systems in their
\revised{ChatGPT, in its current state}, increased \revised{participants'} frustration levels, led to uncertainty, and in some cases induced self-doubt, due to the lack of transparency and clarity in its behavior and communication. This highlights the need for caution; while it provides good answers in straightforward cases, it tends to give incorrect or confusing responses in more complex scenarios. \textit{``For anything that wasn't super standard, ChatGPT struggled to easily give useful answers (PT-1)''}---Expert developers can navigate this, finding value in the AI responses, but novices might struggle or learn incorrect practices.

% These consequences were due to its current faults stemming from HAI guideline violations. 
%\boldification{What we want to do next}
Our findings provide foundational insights for future convo-genAI design towards enhanced human-AI interaction, subsequently informing the current consequences of using such tools for acquiring new knowledge and skills. We foresee that with careful co-design, genAI holds immense potential in helping novices learn software engineering. We plan to use the insights from this study to implement and evaluate genAI-embedded pedagogical tools that foster critical thinking and support learning.